\documentclass[12pt,a4paper]{article}
\usepackage{graphicx,subfigure}
\usepackage{amsmath}
\usepackage{epsf,amsfonts,amssymb,epsfig,color,fancyhdr,lastpage}
\usepackage[a4paper,left=2.5cm,right=2.5cm,top=2.5cm,bottom=2.5cm]{geometry}
\usepackage[colorlinks=true,linktocpage=true,linkcolor=blue,citecolor=blue]{hyperref}

\numberwithin{equation}{section}

\title{\LARGE \bf Localized plasma balls}

\author{Pau Figueras and Saran Tunyasuvunakool}

\date{}

\begin{document}

\maketitle

\thispagestyle{empty}

\begin{center}
DAMTP, Centre for Mathematical Sciences, \\
\vskip .25cm
Wilberforce Road, Cambridge CB3 0WA, U.K.
\\
\vskip .5cm
\texttt{p.figueras@damtp.cam.ac.uk, s.tunyasuvunakool@damtp.cam.ac.uk}
\end{center}

\vskip 1cm
\begin{abstract}
In this paper we numerically construct localized black hole solutions at the IR bottom of the confining geometry of the AdS soliton. These black holes should be thought of as the finite size analogues of the domain wall solutions that have appeared previously in the literature.  From the dual CFT point of view, these black holes correspond to finite size balls of deconfined plasma surrounded by the confining vacuum. The plasma ball solutions are parametrized by the temperature.  For temperatures well above the deconfinement transition, the dual black holes are small and round and they are well-described by the asymptotically flat Schwarzschild solution. On the other hand, as the temperature approaches the deconfinement temperature, the black holes look like pancakes which are extended along the IR bottom of the space-time. On top of these backgrounds, we compute various probes of confinement/deconfinement such as temporal Wilson loops and entanglement entropy. 
\end{abstract}

\newpage
\setcounter{page}{1}

\section{Introduction}
\label{sec:intro}
The AdS/CFT correspondence \cite{Maldacena:1997re,Witten:1998qj,Gubser:1998bc} offers a consistent theoretical framework that allows us to study certain strongly coupled gauge theories from first principles and in a controlled manner. In this paper we shall use this tool to study the finite temperature ({\it i.e.,} the canonical ensemble) equilibrium physics of confining strongly coupled gauge theories in the large $N_c$ limit.

The study of the confinement/deconfinement  phase transition in strongly coupled gauge theories using AdS/CFT was initiated in \cite{Witten:1998zw}. Here we summarize those aspects that we need (see  \cite{Marolf:2013ioa} for a recent extensive review on this and related topics).  
 For a CFT at finite temperature in Minkowski space there cannot be a phase transition at any finite temperature. The reason for this is that by conformal symmetry, any non-zero temperature can be scaled to any other value. Hence, all non-zero temperature states are equivalent. From the bulk point of view, this statement translates into the fact that the planar Schwarzschild--AdS solution is the dominant phase for all temperatures.  One can brake the conformal symmetry by putting the CFT on a sphere. In this case, strictly speaking, only in the $N_c\to\infty$ limit one has a phase transition which has been identified as a confinement/deconfinement phase transition. The bulk dual of this phase transition corresponds to the celebrated Hawking-Page phase transition between thermal AdS and global Schwarzschild--AdS \cite{Hawking:1982dh}.

In this paper we shall be interested in studying confinement/deconfinement in CFTs on flat space with one of the spatial directions compactified on a circle with anti-periodic ({\it i.e.,} Scherk--Schwarz) boundary conditions for the fermions. Having a compact circle breaks conformal invariance and makes it possible to have a non-trivial confinement/deconfinement phase transition as the temperature changes.  From the bulk point of view, the low temperature confining vacuum corresponds to the so called AdS soliton metric \cite{Witten:1998zw,Horowitz:1998ha}. On the other hand, the high temperature deconfined phase corresponds to the planar Schwarzschild--AdS solution. In \S\ref{subsec:preliminaries} we will review in more detail these geometries.

Precisely at the deconfinement temperature, the period of the Euclidean time circle and that of the Scherk--Schwarz circle become equal. Moreover, at this critical temperature the pressure of the plasma vanishes. This suggests that  there might exist a solution of the Einstein equations in the bulk which interpolates between the AdS soliton geometry (confined phase) and the planar Schwarzschild--AdS black hole (deconfined phase). This expectation turns out to be correct and Ref. \cite{Aharony:2005bm} constructed, numerically, such a solution. The existence of such a domain wall solution led the authors of \cite{Aharony:2005bm} to conjecture that finite size black holes, localized at the IR bottom of the AdS-soliton background, should also exist. From the dual CFT perspective, these black holes would correspond to finite size balls of deconfined plasma surrounded by the confining vacuum. In fact, \cite{Aharony:2005bm} argued that finite size plasma balls should generically exist in any confining large $N_c$ gauge theory that exhibits a first order confinement/deconfinement phase transition. In the semiclassical approximation, such plasma balls should be stable. The reason is that in the full quantum theory, plasma balls in confining backgrounds can only evaporate via the emission of colour singlet glueball states. This process is the dual of the Hawking evaporation in the bulk. However, out of the $O(N_c^2)$ degrees of freedom available in the theory, only $O(1)$ correspond to the colour singlet states that can be emitted into the confining vacuum. Therefore, the evaporation process of the plasma balls is suppressed in the large $N_c$ limit \cite{Aharony:2005bm}.

Ref. \cite{Emparan:2009dj} took the first steps towards getting a quantitative understanding of plasma balls in confining backgrounds. This reference considered a particular form of the four-dimensional AdS C-metric \cite{Plebanski:1976gy} corresponding to a black hole moving in an accelerated trajectory inside AdS and cut off the geometry in the IR. This construction can be regarded as the complement of the one in \cite{Emparan:1999wa} to construct localized black holes on the brane in Randall-Sundrum II braneworld models. The solutions of \cite{Emparan:2009dj} confirmed some of the predictions of \cite{Aharony:2005bm}; in particular, the bulk horizon corresponding large plasma balls (or plasma disks in this case) does indeed have a pancake-like shape which is extended along the IR bottom of the geometry.  Also, the properties of the plasma at the interior of the ball (or disk) are those of the deconfined state at the same temperature. However, the solutions of \cite{Emparan:2009dj} also exhibit some peculiarities due to boundary conditions at both the IR and the UV ends of the geometry. For instance, the boundary metric is not flat and it asymptotes to a cone with a certain excess angle. In addition, whilst the radial proper extent of the plasma ball near the deconfinement temperature grows unbounded, it does not imply that the plasma ball extends out to infinity.

In light of these results, it would be desirable get an understanding of the physical properties of these plasma balls in a cleaner setting such as the framework of \cite{Aharony:2005bm}. This is what we do in this paper. By employing numerical methods, we construct localized black holes at the IR bottom of the AdS soliton geometry which are dual to the sought plasma balls. For simplicity, we consider static and spherically symmetric configurations from the point of view of the non-compact directions along the boundary. This reduces the problem to solving non-linear PDEs in two variables.  In this paper we limit our construction to 5 and 6 bulk dimensions. In both cases, the qualitative physics of the plasma balls is the same and hence we do not expect new physics, at least within the same symmetry class, by increasing the number of dimensions. Having found the solutions, we proceed to study some of their physical properties. These black holes are parametrized by their temperature and they exist above the deconfinement phase transition. First, we study the shape of the horizon for plasma balls of different size. We find that small plasma balls are described by approximately spherical black holes in the bulk, whilst large plasma balls are dual pancake-like black holes that extend along the IR bottom of the geometry. Away from the edges of the black hole, we find that indeed the geometry is well-described by a homogeneous black brane at the same temperature. By extracting the stress of the dual field theory, one sees that as we approach the deconfinement temperature, the energy density at the centre of the ball approaches the energy density of a homogeneous black brane at the same temperature. For large plasma balls, the region in the vicity of their edge reduces to a good approximation to the domain wall solution of \cite{Aharony:2005bm}. In particular, we find that the tension is positive.

The rest of this paper is organized as follows. In \S\ref{sec:setup} we explain our set up. In \S\ref{subsec:preliminaries} we review the confinement/deconfinement phase transition in the kind of backgrounds that are relevant for this paper. In \S\ref{subsec:setup} we give the details of our numerical construction of the black holes that are dual to plasma balls. \S\ref{sec:results} devoted to the analysis of some of the physical properties of the plasma balls. In \S\ref{subsec:geometry} we study the horizon geometry of the black holes and in \S\ref{subsec:probes} we analyse various probes of confinement such as temporal Wilson loops and entanglement entropy. In \S\ref{subsec:stresstensor} we present our results for the stress tensor of the dual field theory. In \S\ref{sec:summary} we summarize our results and conclude. We have relegated several aside results to the Appendices. With some minor modifications, one can construct black holes with extended horizons in this kind of confining backgrounds. We call them ``plasma walls" and in Appendix \ref{sec:plasmawalls} we explain how to construct them. Sone of the convergence tests that we have performed are explained in Appendix \ref{sec:convergence}.

\section{Localized plasmaballs: setup}
\label{sec:setup}
In this section we explain our numerical construction of the localized plasmaballs. In \S\ref{subsec:preliminaries} we review the necessary features of both the AdS-soliton metric and the black brane. In \S\ref{subsec:setup} we explain the details of the actual numerical construction.

\subsection{Preliminaries}
\label{subsec:preliminaries}
This subsection provides a brief review of the basics of thermal field theory in Scherk--Schwarz--AdS (SS--AdS) with anti-periodic boundary conditions for the fermions on the SS circle; for more details see the recent review \cite{Marolf:2013ioa} and references therein. 

We are interested solutions to the Einstein vacuum equations in $d+1$ dimensions with SS--AdS boundary conditions, {\it i.e.,} the boundary geometry is (conformal to)  $\mathbb R^{1,d-2}\times \mathbf{S}_\mathrm{SS}^1$. Here the $ \mathbf{S}_\mathrm{SS}^1$ is usually referred to as the Scherk--Schwarz (SS) circle and we will take the fermions in the CFT to be anti-periodic on this circle.\footnote{With this choice of boundary conditions for the fermions we can define a spin structure in the bulk on a manifold in which the SS circle shrinks to zero size. Only with these boundary conditions we can include the AdS-soliton (see below) as an allowed bulk solution \cite{Witten:1998zw}.}  Since we want to use AdS/CFT to study thermal field theories we shall work with the Euclidean section, $t=-\textrm{i}\,\tau$, so that $\tau$ is the Euclidean time and its period $\tau\sim \tau+\beta$ is the inverse temperature. Therefore, in this paper we shall consider $(d+1)$-dimensional asymptotically locally AdS spaces whose boundary metric is $\mathbb R^{d-2}\times\mathbf{S}_\beta\times\mathbf{S}_\mathrm{SS}^1$. Furthermore, since we will only consider static configurations, working in the Euclidean section does not make any practical difference at the level of the numerical construction (see \S\ref{subsec:setup}).

Within this class of geometries, it has been conjectured that the so-called AdS soliton \cite{Witten:1998zw,Horowitz:1998ha} is the actual ground state \cite{Horowitz:1998ha}:
\begin{equation}
ds^2_{soliton}=\frac{\ell^2}{z^2}\left(d\tau^2+\frac{1}{f(z)}\,dz^2+d\mathbf{x}_{d-2}^2+f(z)\,d\theta^2\right)\,,\qquad f(z)=1-\left(\frac{z}{z_0}\right)^d\,,
\label{eqn:soliton}
\end{equation}
where $d\mathbf{x}_{d-2}^2=\sum_{i=1}^{d-2}dx^i\,dx^i$ is the flat metric on $\mathbb R^{d-2}$. The space-time \eqref{eqn:soliton} is smooth and complete iff the period of the SS circle is given by
\begin{equation}
\Delta\theta = \frac{4\,\pi\,z_0}{d}\equiv L\,.
\end{equation}
Near $z=z_0$ the metric \eqref{eqn:soliton} approaches the flat metric on $\mathbb R^{d-2}\times\mathbf S_\beta\times \mathbb R^2$ and therefore the global topology of the space-time is $\mathbf S_\beta\times\mathbb R^d$. This shrinking of the SS circle in the bulk smoothly cuts off the geometry in the IR, which leads to a mass gap and confinement \cite{Witten:1998zw}. Note that in \eqref{eqn:soliton} the period of the Euclidean time circle can be arbitrary and therefore, from the point of view of the canonical ensemble, the AdS-soliton phase \eqref{eqn:soliton} exists at all temperatures. 

By exchanging the thermal and the SS circles, $\tau\leftrightarrow\theta$, in \eqref{eqn:soliton} it is clear that we can immediately write down another solution to the Einstein equations obeying the same boundary conditions:
\begin{equation}
ds^2_{brane}=\frac{\ell^2}{z^2}\left(f_\beta(z)\,d\tau^2+\frac{1}{f_\beta(z)}\,dz^2+d\mathbf{x}^2+d\theta^2\right)\,,\qquad f_\beta(z)=1-\left(\frac{z}{z_h}\right)^d\,,
\label{eqn:brane}
\end{equation}
This is the well-known planar Schwarzschild--AdS black hole. In this solution, the Euclidean time circle is contractible in the bulk and regularity of the geometry \eqref{eqn:brane} at $z=z_h$ fixes the temperature as a function of the IR cutoff:
\begin{equation}
\beta = \frac{4\,\pi\,z_h}{d}\,.
\end{equation}
Note also that in this case SS circle is non-contractible and hence the period of $\theta$ can be arbitrary. 

There is yet another classical solution to the Einstein equations obeying these boundary conditions, namely the pure Poincar\'e--AdS space-time (with both $\tau$ and $\theta$ suitably identified). However, this solution never dominates the thermal ensemble and therefore we will ignore it from now on. 

In this paper we shall be interested in studying the finite temperature phases in SS--AdS. By comparing the thermodynamic quantities, and in particular, the free energy we can determine which is the dominant phase at a given temperature. It turns out that at temperatures $T<\frac{1}{L}$, the AdS-soliton has the lowest free energy and hence it is the dominant phase.\footnote{Note that with the standard counter-terms, thermal Poincar\'e--AdS has zero free energy and the AdS-soliton has a negative free energy which scales with the central charge of the CFT. This contribution can be interpreted as coming from the Casimir effect. We would like to thank R. Emparan and D. Mateos for an enlightening discussion on this. } On the other hand, at temperatures $T>\frac{1}{L}$ is the planar black hole that dominates the thermal ensemble and at $T=T_d=\frac{1}{L}$ there is a first order confinement/deconfinement phase transition.   Given the symmetry between the $\tau$ and $\theta$ circles in the AdS-soliton and the planar black hole geometries, it should not be surprising that the phase transition occurs precisely when the sizes of those two circles become equal. In fact, at $T=T_d$ \eqref{eqn:soliton} and \eqref{eqn:brane} are symmetric under the exchange of the $\tau$ and $\theta$ circles. This led the authors of \cite{Aharony:2005bm} to construct (numerically) a domain wall solution at temperature $T=T_d$ which interpolates between a confined and a deconfined regions. Moreover, the wall has thickness $\sim L$ since at $T=T_d$ the thermal scale and the scale of the mass gap coincide. One can measure the tension of the wall $\mu_{d-1}$ in the field theory and it turns out to be positive. We will come back to this point when we analyse the physics of the plasma balls. 

The existence of this domain wall solution led \cite{Aharony:2005bm} to conjecture there should also exist finite size black holes localized in the IR region of the geometry. From the point of view of the dual CFT, these black holes would correspond to a bubbles of deconfined plasma within the confining vacuum.  More precisely, \cite{Aharony:2005bm} conjectured that generic confining backgrounds should host a one parameter family of black holes which are spherically symmetric in the $p=d-2$ non-compact spatial dimensions, labeled by a mass, which satisfy the following properties:
\begin{enumerate}
\item The radius of the black hole in $p$ dimensions scales with the mass like $m^\frac{1}{p}$.
\item In the interior of these black holes (meaning away from the edge in the $p$ non-compact dimensions) approximates the black brane at $T=T_d$.
\item In the vicinity of the edge, these black holes reduce to a domain wall which interpolates between a black brane at $T_d$ and the AdS soliton.
\end{enumerate}
In this paper we shall provide strong evidence that such black holes do exist by numerically constructing them. Moreover, by extracting their physical quantities we will be able to test some of these conjectures put forward in \cite{Aharony:2005bm}.

As an aside, in Appendix \ref{sec:plasmawalls} we will construct black holes which are extended along the field theory directions. We will call them ``plasma walls". Their thermal properties, at least within the symmetry class of solutions that we consider in this paper, turn out to be qualitatively similar to those of the plasma balls. Therefore, in this paper we shall not discuss them in detail. But it is worth pointing out that these are interesting gravitational solutions in their own right and they could be useful in order to study the dynamics of gravity in SS--AdS. We will leave this for future work.

Before we move on to describe the details of our numerical construction we will outline our conventions. Unless otherwise stated, we will set the AdS radius $\ell$ to one. In order to study the thermal phases in SS--AdS we first recall that by conformal invariance only the dimensionless ratio $\beta/L$ is physical.  It is therefore convenient to fix the scale by setting
\begin{equation}
z_0=1\quad \Rightarrow \quad L = \frac{4\,\pi}{d}\,,
\end{equation}
and then vary the (inverse) temperature $\beta$. Therefore, in our units the deconfinement temperature is $T_d=\frac{d}{4\,\pi}$.

\subsection{Set up}
\label{subsec:setup}

 To construct these black holes we will use the formulation of the Einstein equations proposed in \cite{Headrick:2009pv} and further understood in \cite{Figueras:2011va}. The idea is to solve the vacuum Einstein--DeTurck equations in $d+1$ dimensions with a negative cosmological constant, 
\begin{equation}
R_{ab}+\frac{d}{\ell^2}\,g_{ab} -\nabla_{(a}\xi_{b)} = 0\,,\qquad \xi^a=g^{bc}\left(\Gamma^a_{\phantom a bc}-\bar\Gamma^{a}_{\phantom a bc}\right)\,,
\label{eqn:eoms}
\end{equation} 
 where $\Gamma^a_{\phantom a bc}$ is the Levi-Civita connection compatible with the space-time metric $g$ and $\bar\Gamma^a_{\phantom a bc}$ is the Levi-Civita connection compatible with some reference metric $\bar g$ living on the same space-time manifold $\mathcal M$. For static space-times which are asymptotically either flat, AdS or Kaluza--Klein, and whose boundary conditions are compatible with $\xi^a$ vanishing at the boundaries $\partial\mathcal M$ of the manifold (if any exist), Ref. \cite{Figueras:2011va} showed that all solutions to \eqref{eqn:eoms} are necessarily Einstein. As we shall see below, for our particular problem the boundary conditions that we will impose fall within the class for which the result of \cite{Figueras:2011va} applies, and hence solving \eqref{eqn:eoms} is equivalent to solving the Einstein equations. This method has by now become standard in the field and we refer the reader to \cite{Headrick:2009pv,Figueras:2011va,Wiseman:2011by} for more details.

We seek finite size black holes which are asymptotically SS--AdS, and which are localized in the IR of the geometry. This can be thought of as finite energy (as opposed to finite energy density) excitations about the AdS soliton background.  For simplicity we will only consider static black holes but in view of the results of \cite{Lahiri:2007ae}, it would also be interesting to explore rotating black holes in this backgrounds. We will leave this for future work. 

The AdS soliton is the lowest energy state in this class of metrics \cite{Horowitz:1998ha} ({\it i.e.,} with boundary geometry $\mathbb R^{d-2}\times\mathbf{S}_\beta\times\mathbf{S}_\mathrm{SS}^1$). Since the global topology of the spatial sections of the background is $\mathbb R^d$, one expects that localized black holes in the bulk should exist. These would be static black holes which are asymptotically SS--AdS and with a horizon with $\mathbf S^{d-1}$ spatial topology. Indeed, in the limit in which the black hole is much smaller than the AdS radius and the SS circle radius at infinity, one would expect that such solutions would be well-approximated by the standard asymptotically flat Schwarzschild black hole. As we shall see in \S\ref{subsec:geometry}, this expectation turns out to be correct. Note that from the dual CFT perspective, these black holes correspond to states which are localized on the $\mathbb R^{d-2}$ part of the boundary geometry, and not on the $\mathbf S^1_\mathrm{SS}$ because this is a contractible circle in the bulk. For simplicity we will restrict ourselves to bulk space-times which 
are rotationally symmetric from the point of view of the boundary in the sense that they preserve the $SO(d-2)\times U(1)_\mathrm{SS}$ symmetry of the spatial $\mathbb R^{d-2}\times\mathbf S^1_\mathrm{SS}$ flat boundary metric. From the point of view of the CFT, these black holes should correspond to the plasma balls considered in \cite{Aharony:2005bm}, {\it i.e.,} rotationally symmetric balls of deconfined plasma sitting in the confining vacuum.

\subsubsection{Metric ansatz and boundary conditions}
\label{subsubsec:BCs}
Taking into account our symmetry assumptions, finding a single coordinate system which is adapted to the entire geometry is a rather hard problem in itself. We will overcome this difficulty by using two coordinate charts, one adapted to the asymptotic region far from the black hole and the other one adapted to the near-horizon region. This approach was first used successfully in the numerical construction of 5D localized Kaluza--Klein black holes \cite{Headrick:2009pv}.
 
In the region far from the bulk black hole, the space-time should approach the AdS soliton metric \eqref{eqn:soliton} in a suitable sense.  Since we are interested in preserving the spatial $SO(d-2)\times U(1)_\mathrm{SS}$ symmetry of the boundary metric, we may rewrite \eqref{eqn:soliton}  to make these symmetries manifest as 
\begin{equation}
ds^2_{soliton}= \frac{1}{z^2}\left(d\tau^2+\frac{1}{f(z)}\,dz^2+d\rho^2+\rho^2\,d\Omega_{(d-3)}^2+f(z)\,d\theta^2\right)\,,
\end{equation}
where $f(z)$ has been defined before. For computational purposes, we find it convenient to introduce new compact coordinates $(x,y)$,
\begin{equation}
z=1-y^2\,,\qquad \rho = \frac{k_x\,x}{1-x^2}\,,
\end{equation}
where $k_x$ is a freely adjustable parameter which allows us to stretch the $\rho$ coordinate. With these co-ordinates we can  bring the asymptotic region $\rho\to\infty$ to a finite coordinate distance $x=1$.  In terms of these new coordinates, the AdS soliton metric becomes,
\begin{equation}
ds^2_{soliton}=\frac{1}{(1-y^2)}\left(d\tau^2+\frac{4\,z_0^2}{\bar f(y)}\,dy^2+\frac{k_x^2(1+x^2)^2}{(1-x^2)^4}\,dx^2+\frac{k_x^2\,x^2}{(1-x^2)^2}\,d\Omega_{(d-3)}^2+\bar f(y)\,d\theta^2\right)\,,
\label{eqn:solitonnew}
\end{equation} 
where we have defined $\bar f(y)$ via $f(y)=y^2\,\bar f(y)$.  We are now ready to write down the ansatz for the metric in the far region:
\begin{equation}
\begin{aligned}
ds^2_{Far}=&~\frac{1}{(1-y^2)^2}\Bigg(T\,d\tau^2+\frac{4\,y^2\,\bar f(y)\,S}{d^2}\,d\theta^2+\frac{k_x^2\,x^2\,R}{(1-x^2)^2}\,d\Omega_{(d-3)}^2\\
&\hspace{2cm}+\frac{k_x^2(1+x^2)^2\,A}{(1-x^2)^4}\,dx^2+\frac{4\,B}{\bar f(y)}\,dy^2-\frac{2\,k_x\,(1+x^2)\,F}{(1-x^2)^2(1-y^2)}\,dx\,dy\Bigg)\,.
\end{aligned}
\label{eqn:far}
\end{equation}
Here the functions $\{T,~R,~S,~A,~B,~F\}$ are our unknowns and they depend on both $x$ and $y$. In these coordinates, $x=0$ and $y=0$ correspond to the fix point set of the $SO(d-2)$ symmetry and of the $U(1)_\mathrm{SS}$ symmetry respectively. Regularity there requires that all functions be Neumann except for $F$, which has to satisfy a Dirichlet boundary condition there.  To avoid conical singularities at $x=0$ we must also impose $A=R$; similarly, the avoidance of conical singularities at $y=0$ requires $B=S$ there. The boundary $x=1$ is an asymptotic end, where the space-time should approach the AdS soliton metric \eqref{eqn:solitonnew}. Hence we impose Dirichlet boundary conditions there: $T=R=S=A=B=1$, $F=0$. Similarly, $y=1$ is the boundary of AdS and we impose Dirichlet boundary conditions as before. Recall that by varying the norm of the Euclidean time circle at the boundary we can change the temperature of the bulk black hole. We will make use of this in our construction to efficiently explore the branch of solutions.  Finally, note that \eqref{eqn:far} is not adapted to describe the horizon in the sense that in these $(x,y)$ coordinates the horizon does not lie at a constant value of either of them.  Therefore, in order to avoid unnecessary difficulties when dealing with horizons in non-adapted coordinates, we will ensure that the domain covered by \eqref{eqn:far} does not contain horizons. We will return to this point later when we describe our computational domain.

In the region near the horizon, the space-time should approach the near-horizon region of a topologically spherical black hole. We expect that when the plasma balls are small, the bulk black holes should be a approximately like asymptotically flat static black holes, and hence spherically symmetric in the full $(d+1)$-dimensional sense. On the other hand, as the plasma balls become larger, the presence of a non-zero cosmological and the non-trivial topology of the IR bottom of the geometry will become important, and the black hole should be highly deformed from perfect spherical symmetry. Therefore, it is desirable to have an ansatz which can be adapted to suit each of these two different regimes. To achieve this, we write the near-horizon ansatz as
\begin{equation}
\begin{aligned}
ds^2_{Near} =&~\frac{1}{(1-r^2)^2}\bigg[r^2\,g(r)\,T'\,d\tau^2 + r_0^2\Big((1-a^2)^2\,R'\,d\Omega_{(d-3)}^2+a^2(2-a^2)\,S'\,d\theta^2\Big) \\
&\hspace{2cm}
+\frac{4\,r_0^2\,A'}{g(r)}\,dr^2 + \frac{4\,r_0^2\,B'}{2-a^2}\,da^2 + 2\,r\,F'\,dr\,da\bigg]\,.
\end{aligned}
\label{eqn:near}
\end{equation} 
where $g(r)$ is a freely specifiable function.   The functions $\{T',~R',~S',~A',~B',~F'\}$ are our unknowns. The radial coordinate $r$ in \eqref{eqn:near} ranges from $r=0$ (horizon)  up to some $r=r_{out}$ that we can,  in principle,  freely choose. In practice, $r_{out}$ can neither be too large nor too small, and we find that $r_{out}\simeq 0.6$ works well. In our calculations we will choose $g(r)$  so that the near-horizon geometry \eqref{eqn:near} approaches that of the asymptotically AdS Schwarzschild black hole in the corresponding number of space-time dimensions. Regularity at the horizon requires that all functions are Neumann there. On top, we have to impose that $T'=A'$ to ensure that the space-time metric and the reference metric have the same surface gravity. Here $r_0$ is a dimensionless parameter  and it is related to the surface gravity as
\begin{equation}
\kappa^2  =\frac{g(0)^2}{4\,r_0^2}\,. 
\label{eqn:kappa}
\end{equation}
$a$ is an angular co-ordinate, whose range is $0\leq a \leq 1$.  At $a=0$  the SS circle shrinks to zero and regularity requires that all functions are Neumann there except for $F'$, which has to vanish; to avoid conical singularities we impose $B'=S'$. Similarly, at $a=1$ it is the $(d-3)$-sphere that shrinks to zero size;  regularity requires that all functions satisfy a Neumann boundary condition there and $F'=0$. Absence of conical singularities imposes $B'=R'$ at $a=1$.

\subsubsection{Computational domain and reference metric}

In our construction, the far region coordinates $(x,y)$  and the near region coordinates $(r,a)$ are simply related by
\begin{equation}
x = r\,(1-a^2)\,,\qquad y = r\,a\,\sqrt{2-a^2} \,.
\end{equation}
Note that by construction the range of the far region coordinates is $0\leq \{ x\,,y \} \leq 1$. This condition limits the range of $r$ in the inner region patch. In terms of the far region $(x,y)$ coordinates, the horizon $(r=0)$ appears point-like, but this is just a coordinate artefact.  

\begin{figure}[t]
\begin{center}
\includegraphics[scale=0.6]{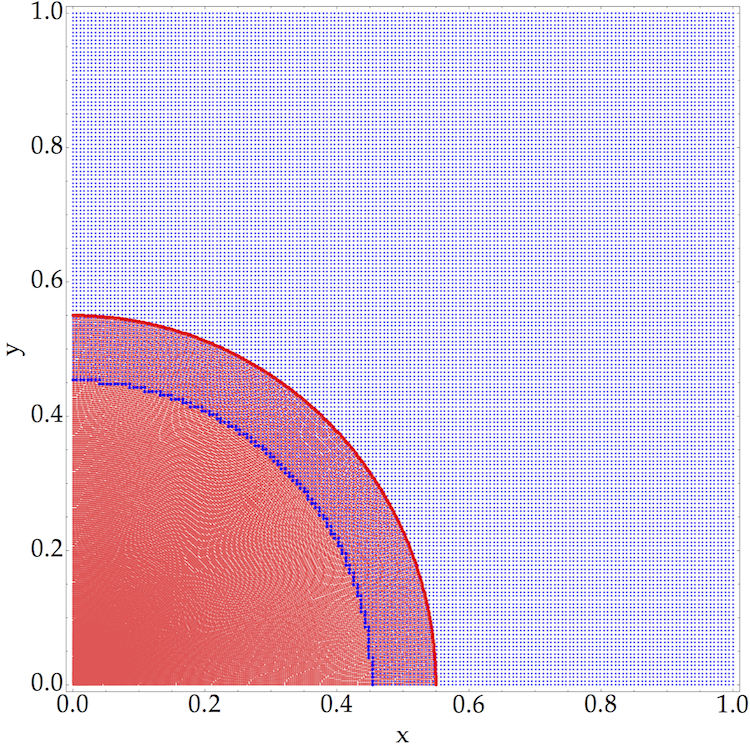}
\end{center}
\caption{Computational domain. The blue dots correspond to the grid points in the far region patch and the red dots are the grid points in the near region patch expressed in the far region coordinates. The thicker red (blue) points indicate the boundary points whose values are obtained by interpolation of the metric functions in the far (near) patch.}
\label{fig:domain}
\end{figure}

We solve \eqref{eqn:eoms} numerically using sixth-order finite differences and Newton's method. One advantage of using finite differences is that we can cover our domain in two simple patches.  In Fig. \ref{fig:domain} we display our computational domain, which is covered by two co-ordinate patches, the far patch (in blue) and the near patch (in red). As this figure shows, the two coordinate charts overlap in a certain region; by appealing to the uniqueness of the solution in the whole domain, we can use this overlapping region to transfer the information between the values of the metric functions in the two patches. Indeed, the values of the functions at the boundary of the overlapping region in a certain patch (thicker dots in Fig. \ref{fig:domain}) are obtained by interpolating the values of the functions in the other patch.  The order of the interpolation that we use is consistent with the order of the differentiation, which guarantees that the interpolation does not spoil the global convergence properties of our numerical solutions.  Even though in principle the size of the overlapping region could be arbitrary, in the actual implementation we found that having a relatively small overlapping between the patches, as in Fig. \ref{fig:domain}, worked better. 

In order to solve \eqref{eqn:eoms} we need to provide a global reference metric in the space-time manifold as part the gauge fixing procedure. Following \cite{Headrick:2009pv}, we can achieve this by a simple interpolation between the near-horizon geometry and the AdS soliton metric \eqref{eqn:solitonnew}:
\begin{equation}
d\bar s^2=\big[\big(1-I(r;d_\textrm{min},d_\textrm{max})\big)\,\bar g^{Near}_{ab} + I(r;d_\textrm{min},d_\textrm{max}))\,\bar g^{Far}_{ab} \,\big]\,dx^a\,dx^b\,,
\label{eqn:refmetric}
\end{equation}
where we have used the interpolation function
\begin{equation}
I(r;d_\textrm{min},d_\textrm{max}) = \left\{
\begin{array}{cl}
0\, &,\,r\leq d_\textrm{min} \\
\frac{1}{2}-\frac{1}{2}\tanh\Big[\cot\Big(\pi\big(1+\frac{d_\textrm{max}-r}{d_\textrm{max}-d_\textrm{min}}\big)\Big)\Big] &,\,d_\textrm{min}<r<d_\textrm{max} \\
1 &,\,d_\textrm{max}\leq r
\end{array}
\right.\,
\label{eqn:interp}
\end{equation}
 and $d_\textrm{min}$, $d_\textrm{max}$ are freely adjustable parameters. These parameters control the size of the overlapping region between the near-horizon geometry and the AdS-soliton metric. In our calculations we typically used $d_\textrm{min}=0.2$ and $d_\textrm{max}=0.9$. Note that \eqref{eqn:interp} is a smooth compactly supported function; this is useful to ensure that the reference metric \eqref{eqn:refmetric}  satisfies our boundary conditions in all regions.  In \eqref{eqn:refmetric}, $\bar g^{Far}$ denotes the  AdS soliton metric, which can be obtained from \eqref{eqn:far} by setting $T=R=S=A=B=1$, $F=0$; similarly, $\bar g^{Near}$ denotes the near-horizon black hole metric and it is obtained from \eqref{eqn:near} by setting $T'=R'=S'=A'=B'=1$, $F'=0$.
 
As \cite{Aharony:2005bm} conjecture, and we confirm in this paper, there exists a unique branch of plasma balls labelled by the temperature. Therefore, we can explore the branch of solutions by varying the parameter $r_0$ in \eqref{eqn:near}, since it controls the surface gravity $\kappa$ of the black hole.  Note, however, that if in  \eqref{eqn:near} we use the function $g(r)$ corresponding to the global Schwarzschild--AdS black hole, by varying $r_0$ we will only be able to explore a finite range of temperatures that may not include the deconfinement temperature.\footnote{Recall that there is a minimum temperature along the global Schwarzschild--AdS black hole branch of solutions and in general this minimum temperature will not coincide with the deconfinement temperature.}   We can overcome this difficulty by introducing another parameter corresponding to the value of $T$ at the boundary, $T\big|_{y=1}=T_0$. Changing $T_0$ corresponds to changing the norm of the Euclidean time circle at the boundary, which in turn implies that the inverse temperature is modified as 
 \begin{equation}
 \beta = \frac{2\,\pi\,\sqrt{T_0}}{\kappa}\,.
 \end{equation}
Note that we also have to set $T=T_0$ on the AdS soliton end of the geometry, $x=1$.
 
Finally we mention that we have checked convergence by analysing the numerical errors in the solution at different resolutions (see Appendix \S\ref{sec:convergence}  for more details) and we have good evidence that indeed our numerical solutions approximate a continuum solution. For the data presented in the next section, we have used 73852 grid points (equivalent to a resolution of $175 \times 275$ in the inner patch and $175 \times 175$ in the outer patch). The error in our numerical solution can be estimated by considering the maximum magnitude of the scalar $1 + R / d (d+1)$ on the grid, which suggests a relative error between 0.0001\% and 0.01\% for medium-sized plasma balls. Another useful measure of error is the maximum norm of the DeTurck vector $\sqrt{\xi^a \xi_a}$, which is also found to the have about the same order of magnitude as the Ricci scalar above.
The highly deformed geometry of larger plasma balls naturally induces larger errors, and the numerical code typically fails when the error is estimated to be of order 1\%.

\section{Results}
\label{sec:results}
 
 In this section we present our results.  In \S\ref{subsec:geometry} we will characterize the geometry of the plasma balls  along the branch of solutions. In \S\ref{subsec:probes} we study various probes of confinement/deconfinement in the background of the plasma balls. In \S\ref{subsec:loops} we study temporal Wilson loops  and in \S\ref{subsec:entropy} we study the entanglement entropy.  Finally, in \S\ref{subsec:stresstensor} we provide our results for the stress tensor of the dual field theory.

\subsection{Characterizing the geometry of plasma balls}
\label{subsec:geometry}
In this subsection we characterize geometry of the horizon of the plasma balls. For small plasma balls ({\it i.e,} high temperatures), the dual black holes are much smaller than the radius of AdS and the size of the SS circle. Hence they should approximately be like the asymptotically flat Schwarzschild solution and the geometry of the spatial cross-sections of the horizon should be that of the round sphere. In the opposite limit, when the temperature approaches the deconfinement temperature, the plasma balls should approach the domain wall solutions of \cite{Aharony:2005bm}. Hence we expect that the horizon will have a finite extent in the direction orthogonal to the IR bottom whilst its typical size along the IR floor will diverge. Therefore, for large plasma balls, the horizon of the dual black holes should look like a pancake which extends along the IR bottom of the space-time.

\begin{figure}[t!]
\begin{center}
\includegraphics[width=15cm]{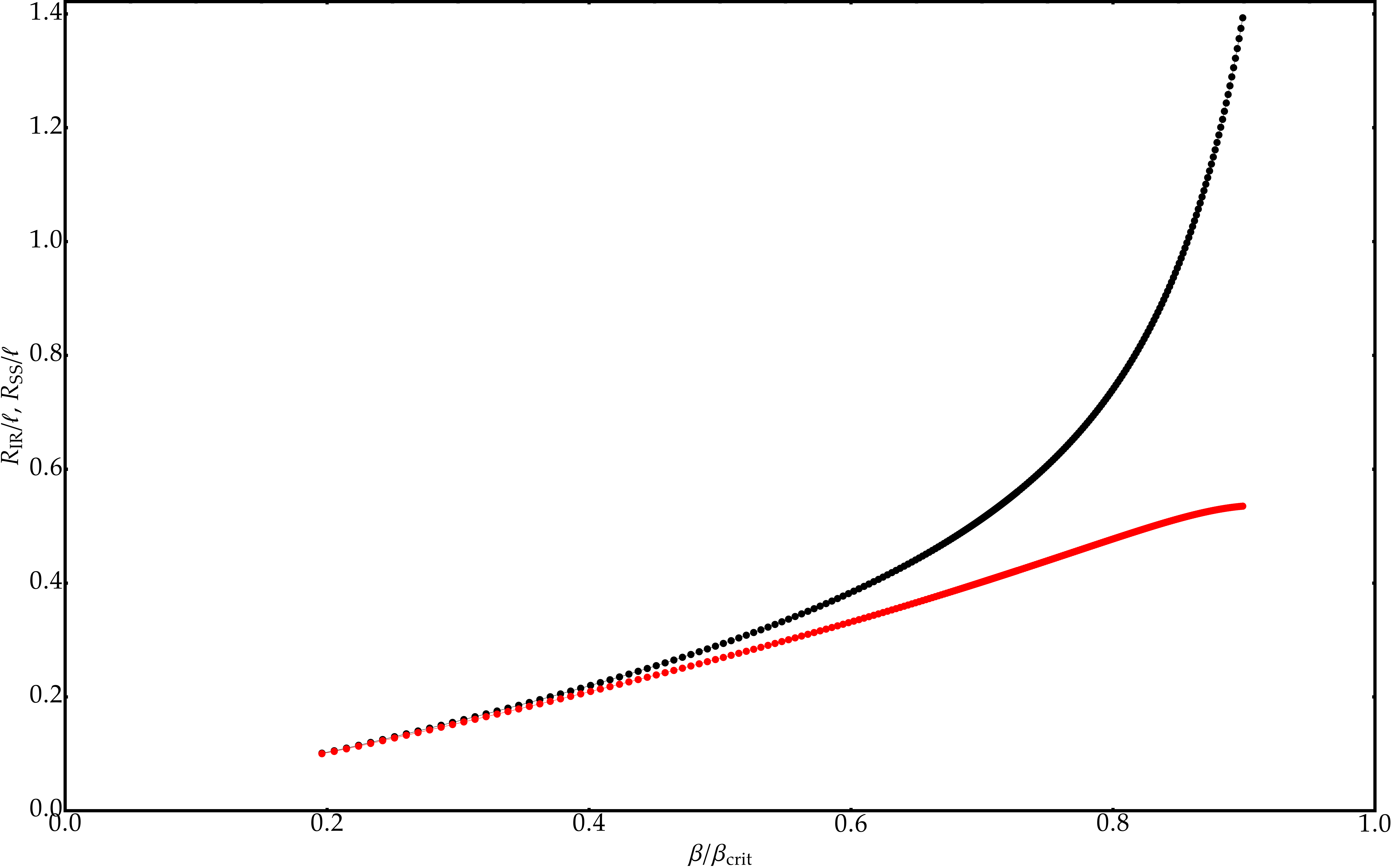}
\end{center}
\caption{Plasma ball horizons can be characterized by two quantities: the radial extent $R_\mathrm{IR}$ (black) of the spatial $S^2$ along the IR bottom and the radius $R_\mathrm{SS}$ (red) of the SS circle at the origin of the non-compact spatial dimensions. Here we plot these two quantities as functions of the scaled inverse temperature.}
\label{fig:radii}
\end{figure}

To study the actual shape of the horizon, we can consider the induced metric on the spatial cross sections of the horizon:
\begin{equation}
ds_H^2=r_0^2\left((1-a^2)^2\,R'\,d\Omega_{(d-3)}^2+a^2(2-a^2)\,S'\,d\theta^2 + \frac{4\,B'}{2-a^2}\,da^2\right)\,,
\label{eqn:inducedH}
\end{equation}
where the various functions are evaluated at $r=0$ (horizon). Recall that the topology of the horizon is $\mathbf S^{d-2}$; we can measure the deformation of the horizon sphere by comparing the size of the SS circle at the $a=1$ equator and the size of the round $\mathbf S^{d-3}$ at the other equator, $a=0$:
\begin{equation}
R_\mathrm{SS} = r_0 \, \sqrt{S'}\big|_{r=0,a=1}\,,\qquad R_\mathrm{IR} = r_0\,\sqrt{R'}\big|_{r=0,a=0}\,.
\end{equation}

For a round sphere, these two radii should be approximately equal. On the other hand, for a pancaked black hole we should have $R_\mathrm{IR}\gg R_\mathrm{SS}$. In Fig. \ref{fig:radii} we display these radii, measured in units of the AdS radius, as a function of the inverse temperature. As this plot shows, in the high temperature limit both radii are equal, to a very good approximation, and the black hole in the bulk is small compared to the radius of AdS. This shows that indeed, the geometry of the horizon is that of an approximately round sphere. As we decrease the temperature towards the deconfinement temperature, we see that $R_\mathrm{SS}$ tends to saturate at a value around 0.535. This suggests that in the limit $T\to T_d$, the extent of the horizon in the orthogonal direction to the IR bottom is finite.  On the other hand, the size of the horizon along the IR bottom of the geometry is measured by $R_\mathrm{IR}$ and this quantity diverges at the deconfinement temperature. Of course, our expectation is that the plasma ball should approach the black brane solution as $T \rightarrow T_d$, where the SS circle has a constant size everywhere. With our choice of parameters, this means that $R_\mathrm{SS}$ should approach the value of $1/2$ as $T \rightarrow T_d$, and thus we conjecture that the apparent saturation in the value of $R_\mathrm{SS}$ is actually a turning point beyond which it will decrease to $1/2$.

\begin{figure}[t]
\begin{center}
\includegraphics[width=15cm]{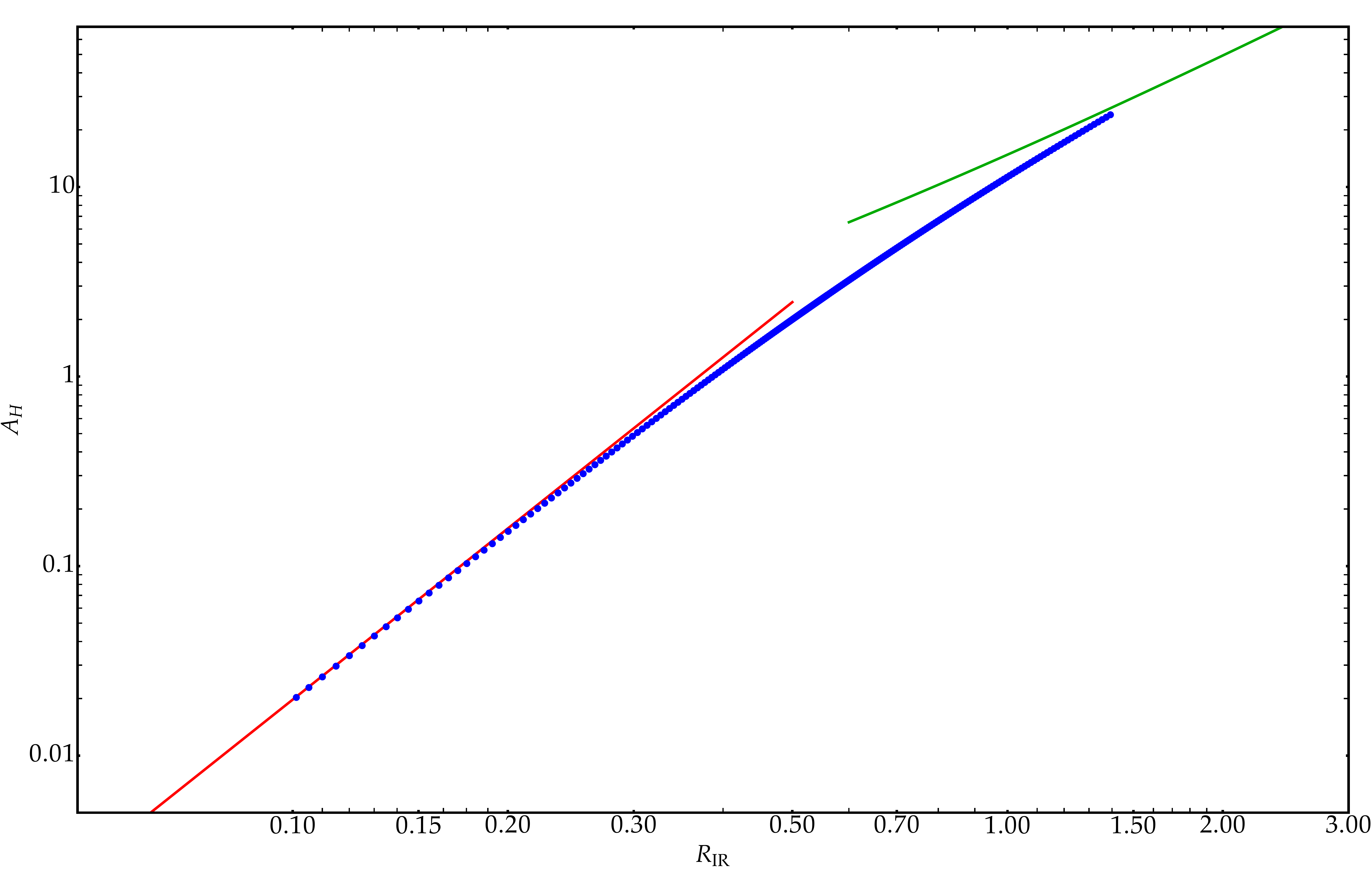}
\end{center}
\caption{Horizon area $A_H$ as a function of its radial extent $R_\mathrm{IR}$ along the IR bottom for 5D plasma balls. The red line shows the function $A_H = 2 \pi^2 R_\mathrm{IR}^2$, which is the scaling behaviour of a perfectly-$S^3$ horizon. The green line shows the function $A_H = 2 \pi^2 \left( R_\mathrm{IR}^2 R_\mathrm{SS} + R_\mathrm{IR} R_\mathrm{SS}^2 \right)$, which is the scaling behaviour of a horizon which is only stretching in the spatial $S^2$ direction while the SS circle maintains a constant size $R_\mathrm{SS} \equiv 1/2$. As $R_\mathrm{IR} \rightarrow \infty$ the behaviour becomes effectively $A_H \sim R_\mathrm{IR}^2$.}
\label{fig:radiuslog}
\end{figure}

In Fig. \ref{fig:radiuslog} we plot the horizon area, ${\mathcal A}_H$, as a function of its radial extent $R_\mathrm{IR}$ along the IR bottom for the 5D plasma balls. Notice the logarithmic scale in $x$-axis of this plot. For small values of $R_\mathrm{IR}$ the area of the horizon scales like $R_\mathrm{IR}^3$, which is the expected behaviour for a spherically symmetric 5D static Schwarzschild black hole. On the other hand, for large plasma balls, the horizon area scales like $R_\mathrm{IR}^2$. This is the behaviour for a 5D black brane whose horizon is infinitely extended in two directions whilst the third one is compact. Therefore, our results indicate that even though the plasma balls that we have constructed are not parametrically much larger than the radius of AdS, they already exhibit some of the expected behaviour of a black brane. In fact, the Smarr relation in 5D implies that the mass of the large plasma balls scales with the radius like $R_\mathrm{IR}^2$, just as the horizon area. Therefore,  our results confirm the predictions of \cite{Aharony:2005bm}.  We will provide additional evidence from the analysis of the stress tensor of the dual CFT.

Finally, in order to get a better intuition about the actual geometry of the horizon, we embed the horizon geometry, \eqref{eqn:inducedH}, into  Euclidean $\mathbb R^{d-1}\times \mathbf S^1_\mathrm{SS}$ space,
\begin{equation}
ds^2_E=dX^2+dY^2 + X^2\,d\Omega_{(d-3)}^2 + Y^2\,d\theta^2\,.
\end{equation}

\begin{figure}[t]
\begin{center}
\includegraphics[width=15cm]{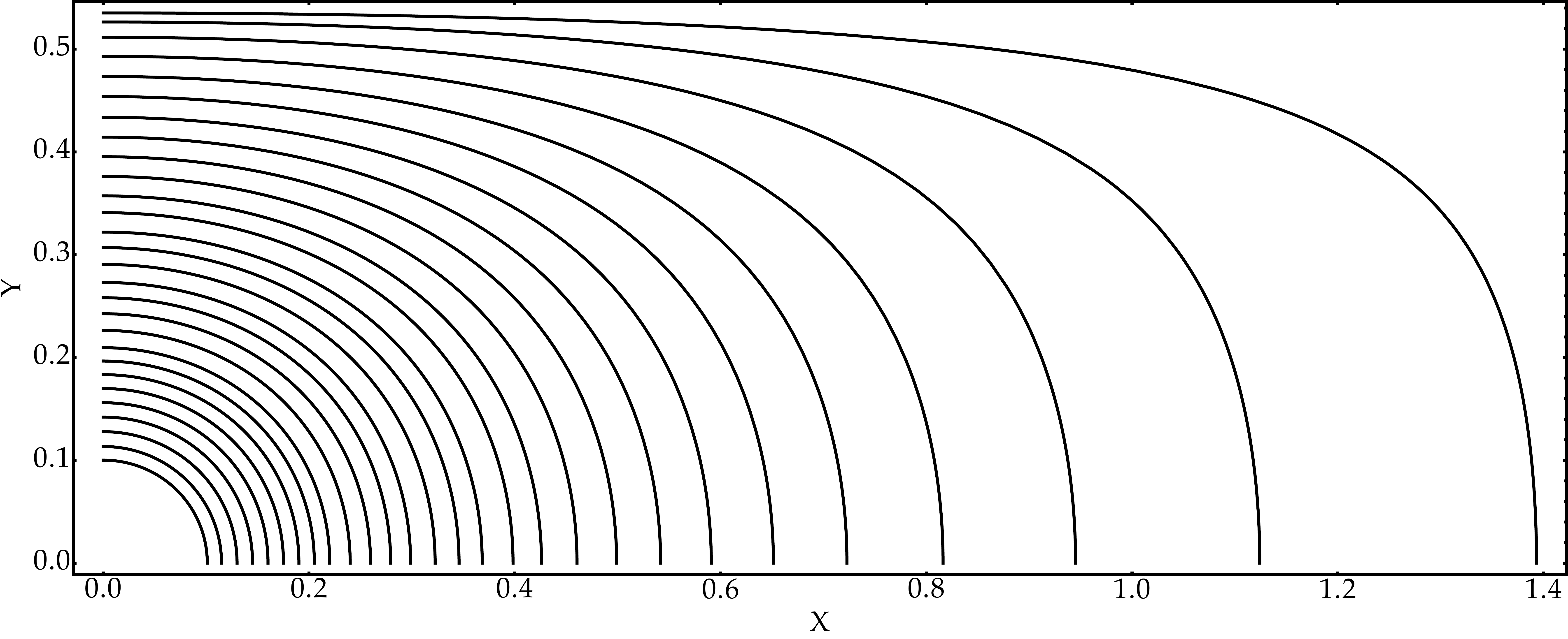}
\end{center}
\caption{Embeddings of the horizon geometry into Euclidean $\mathbb R^{d-1} \times \mathbf S^1_\mathrm{SS}$ for solutions at $\beta / \beta_\mathrm{crit}$ roughly equally spaced between 0.2 and 0.9.}
\label{fig:embed}
\end{figure}

In Fig. \ref{fig:embed} we plot $Y$ vs. $X$ for some of the solutions that we have constructed. As this plot shows, small plasma balls appear to be round, whilst big plasma balls have a pancake-like shape, thus confirming the previous expectations.

\subsection{Probes of confinement}
\label{subsec:probes}
In this subsection we investigate the behaviour of various probes of confinement in the geometries of the plasma balls. In \S\ref{subsec:loops} we analyse temporal Wilson loops, and in \S\ref{subsec:entropy} we consider the entanglement entropy of certain spherically symmetric regions.

\subsubsection{Wilson loops}
\label{subsec:loops}

The standard order parameter for  deconfinement is the expectation value of a Wilson loop wrapping the Euclidean time circle:
\begin{equation}
\langle |\textrm{Tr}(W)|\rangle =\left\langle\left| \frac{1}{N}\,\textrm{Tr}\left(P\,e^{i\oint_{\mathcal C} A_\tau\tau}\right)\right|\right\rangle\,.
\end{equation}
The temporal Wilson loop measures the cost in free energy of perturbing the system by an external quark. If the system is in a confining state, then the cost in free energy is infinite and $\langle |\textrm{Tr}(W)|\rangle =0$; in a deconfined state, the cost in free energy is finite and hence the temporal Wilson loop has a finite expectation value \cite{Witten:1998zw}.

Refs. \cite{Maldacena:1998im,Rey:1998ik} have provided a prescription to calculate Wilson loops in AdS/CFT. According to this prescription, on should consider the action of a classical string wrapping the Euclidean time circle on the boundary in a contour ${\mathcal C} = P\times \mathbf S^1_\beta$, where $P$ is a point in the transverse directions,  and smoothly extending into the bulk,
\begin{equation}
\langle |\textrm{Tr}(W)|\rangle  \sim e^{-S_{string}}\,.
\end{equation}
This classical string action ({\it i.e.,} the area of a minimal surface) diverges and its divergence is proportional to the length of circumference $\mathcal C$ on the boundary. Therefore, we can define a regularized classical string action by subtracting this universal infinite piece, and this is what we can meaningfully compare with the CFT predictions. It is this regularized area that we shall calculate below. 

\begin{figure}[t]
\begin{center}
\includegraphics[scale=0.7]{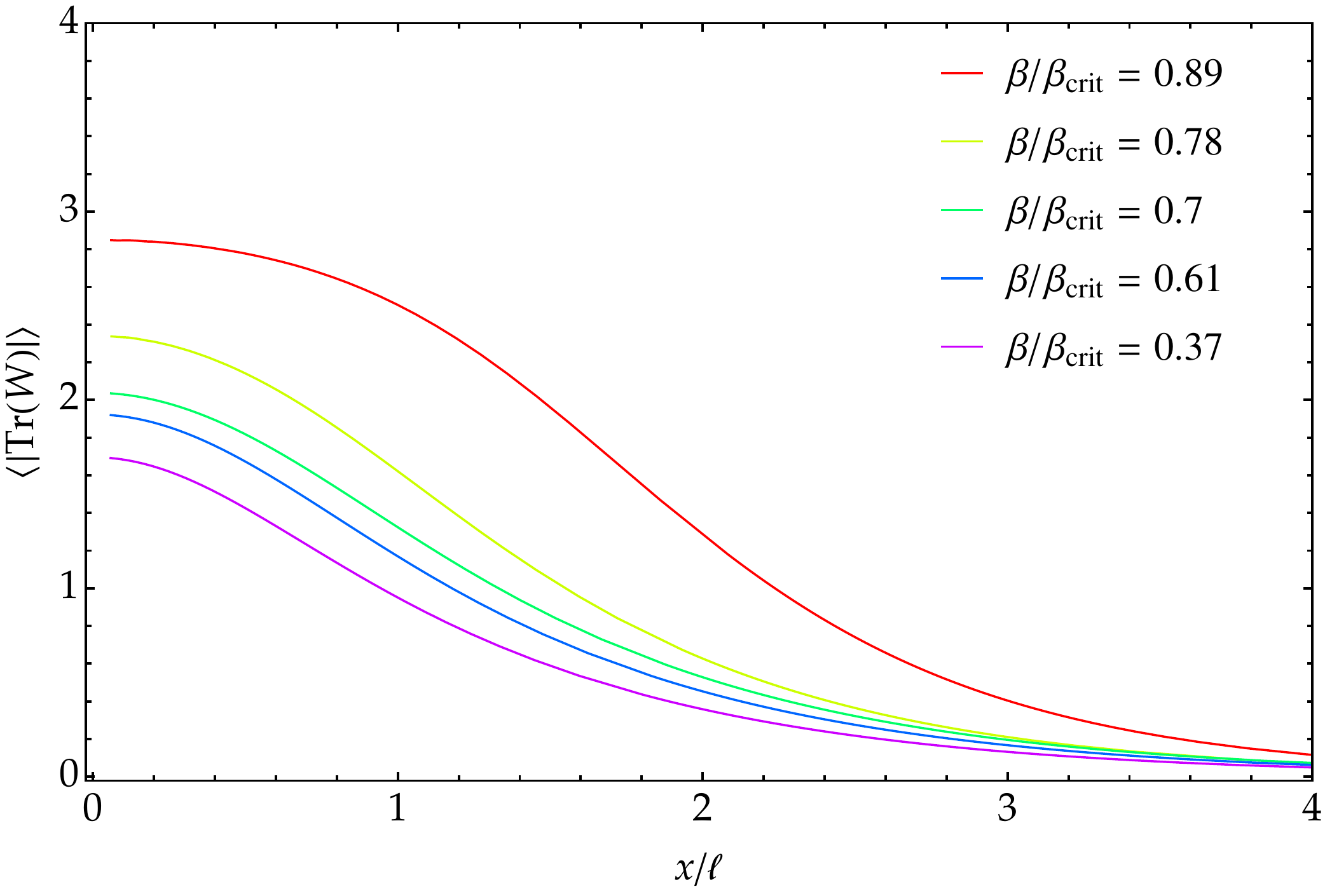}
\end{center}
\caption{Expectation value of temporal Wilson loops, $\langle |\textrm{Tr}(W)|\rangle$, for different temperatures. For temperatures closer to the deconfinement temperature, the black holes in the bulk are larger and hence $\langle |\textrm{Tr}(W)|\rangle$ is non-zero over a larger region on the boundary. The size of this region provides an estimate of the actual size of the plasma ball.}
\label{fig:wilson}
\end{figure}

When the Euclidean time circle is non-contractible in the bulk, then no classical string world-sheet ending on $\mathcal C$ exists and hence $\langle |\textrm{Tr}(W)|\rangle\simeq 0$ \cite{Witten:1998zw}. This is the expected result for the confining state, whose bulk dual is the AdS soliton. On the other hand, if the Euclidean time circle is contractible, then a minimal surface ending on  $\mathcal C$ and which smoothly caps off in the bulk does exist and hence $\langle |\textrm{Tr}(W)|\rangle\neq 0$. This is the deconfined phase and its dual geometry is the black brane.   For our plasma ball geometries, the Euclidean time circle is contractible in the bulk because, at any finite temperature, there is a finite size horizon sitting at the IR bottom of the geometry. Therefore, we expect that a minimal surface ending on $\mathcal C$ should exist and hence the temporal Wilson loop should have a non-zero expectation value. Indeed, from the boundary CFT point of view, there is a finite region of deconfined plasma and the temporal Wilson loop should be sensitive to it. Since $\langle |\textrm{Tr}(W)|\rangle$ is a probe of deconfinement, the expectation value of this operator should provide a measure of the size of the deconfined region. In other words, if the size of the Wilson loop on the boundary is much larger than the size of the deconfined region, then the expectation value should be exponentially small; on the other hand, if the size of the Wilson loop is of the same order as the size of the deconfined region, then the Wilson loop should have a finite expectation value.

To see this, in this paper we consider temporal Wilson loops that end at some fixed value of the radial coordinate $x$ along the boundary \eqref{eqn:far}. Therefore, for our temporal Wilson loops, $\mathcal C=x\times\mathbf S^1_{\beta}$. In Fig. \ref{fig:wilson} we display $\langle |\textrm{Tr}(W)|\rangle$ as a function of the radial co-ordinate on the boundary, $x$, for plasma balls of different sizes ({\it i.e.,} different temperatures). As this figure shows, when the size $x$ of the Wilson loop on the boundary is comparable to the size of the bulk horizon along the IR bottom, the expectation value of the Wilson loop is non-zero. In fact, the typical size of plasma ball can be estimated by the value of $x$ for which  $\langle |\textrm{Tr}(W)|\rangle$ has decayed by one $e$-fold with respect to the corresponding value
at the centre of the ball. The value that we get using this simple estimate is comparable to the size of the black hole along the IR bottom, $R_\mathrm{IR}$. Therefore, we conclude that the expectation value of the temporal Wilson loops offers a measure of the size of the deconfined region on the boundary. 

\subsubsection{Entanglement entropy}
\label{subsec:entropy}

Refs. \cite{Ryu:2006bv,Ryu:2006ef} proposed a prescription to compute the entanglement entropy in AdS/CFT for large $N_c$ conformal field theories with AdS$_{d+1}$ gravity duals. According to the this prescription, the entanglement entropy between regions $A$ and $B$ of the boundary CFT is given by the area of a $(d-1)$-dimensional minimal surface $\gamma_A$ in the bulk which ends at boundary of the region $A$, $\partial A$:
\begin{equation}
S_A=\frac{\textrm{Area}(\gamma_A)}{4\,G_{N}^{d+1}}\,.
\label{eqn:entanglement}
\end{equation}
This expression is divergent, but the divergent pieces only depend on the size of $A$ along the boundary directions. In fact, the leading order divergence is proportional to the area of the boundary $\partial A$.  It is the universal finite piece of $S_A$ that can be meaningfully compared with the dual field theory calculations. 

Subsequently the proposal of \cite{Ryu:2006bv,Ryu:2006ef} has been generalized to time-dependent situations \cite{Hubeny:2007xt} and, more relevant for this paper,  to confining backgrounds \cite{Klebanov:2007ws} (see also \cite{Nishioka:2006gr}). In this reference, the authors considered the entanglement between a strip of width $l$ and its complement and found a phase transition as a function of $l$: for $l<l_\textrm{crit}$ the entanglement entropy is given by the area of a connected surface in the bulk, whilst for $l>l_\textrm{crit}$ it is a disconnected surface that dominates. In the former case, the (finite part of the) entanglement entropy is $O(N_c^2)$ and in the latter is of $O(1)$.  This phase transition is qualitatively similar to the confinement/deconfinement phase transition at finite temperature. Furthermore, \cite{Klebanov:2007ws} argued that the existence of these two kinds of surfaces should be a generic feature of confining backgrounds. Hence, it seems plausible that the entanglement entropy can be used as a probe of confinement in gauge theories with gravity duals. 

The plasma ball backgrounds that we have constructed offer a very non-trivial background where we can use entanglement entropy to probe confinement/deconfinement. Note that our backgrounds are different from those considered by \cite{Ryu:2006bv,Ryu:2006ef} in a crucial way: because the black hole in the bulk is of finite size, in the dual CFT there is a finite region of deconfined plasma surrounded by the confining vacuum. Therefore, our backgrounds are finite energy excitations above the confining vacuum. We shall proceed as follows. Using the prescription of \cite{Ryu:2006bv,Ryu:2006ef}, for a given plasma ball background at some temperature $T>T_d$, we compute the quantum entanglement between a region $A=P\times\mathbf{S}^{d-3}\times\mathbf{S}^1_\mathrm{SS}$ and its complement.  Here $P$ denotes a point $x$ in the radial direction along the non-compact boundary directions. Hence, the regions that we consider in this paper are spherically symmetric from the boundary point of view. 

At a given $T>T_d$ the plasma balls are of finite size. Therefore we would expect that the entanglement entropy is sensitive to the size of the ball because most of the contributions to the entanglement entropy come from the correlations near the boundary of the entangling region. Therefore,  when the size $R_{EE}$ of the boundary $\mathbf S^{d-3}$ is less than the size of the of ball of deconfined plasma, $R_{ball}$, we will be probing entanglement between the $O(N_c^2)$ degrees of freedom of the deconfined plasma. In this case,  we would expect that the bulk minimal surface should look like a spherically symmetric surface in the background of a black brane. By symmetry, such surface should cap off smoothly in the bulk at some finite $z$ above the horizon of the black brane  \eqref{eqn:brane}. On the other hand, for $R_{EE}>R_{ball}$ we will be probing entanglement in the confining vacuum. In this situation, the minimal surface should be similar to a minimal surface in the AdS soliton background, which should close off smoothly at the IR bottom of the geometry (at least for sufficiently large radius $R_{EE}$). As we will see below, our results show $S_A$ exhibits a phase transition: for $R_{EE}<R_{crit}$ the minimal surfaces have a shrinking $\mathbf S^{d-3}$ in the bulk whilst the $\mathbf S^1_\mathrm{SS}$ is always of finite size. On the other hand,  for $R_{EE}>R_{crit}$ the minimal surface has an $\mathbf S^1_\mathrm{SS}$ which shrinks to zero size at the IR bottom of the geometry and the $\mathbf S^{d-3}$ is everywhere finite. This phase transition is somewhat similar to the one observed in \cite{Klebanov:2007ws}. 

\begin{figure}[t]
\begin{center}
\includegraphics[scale=0.7]{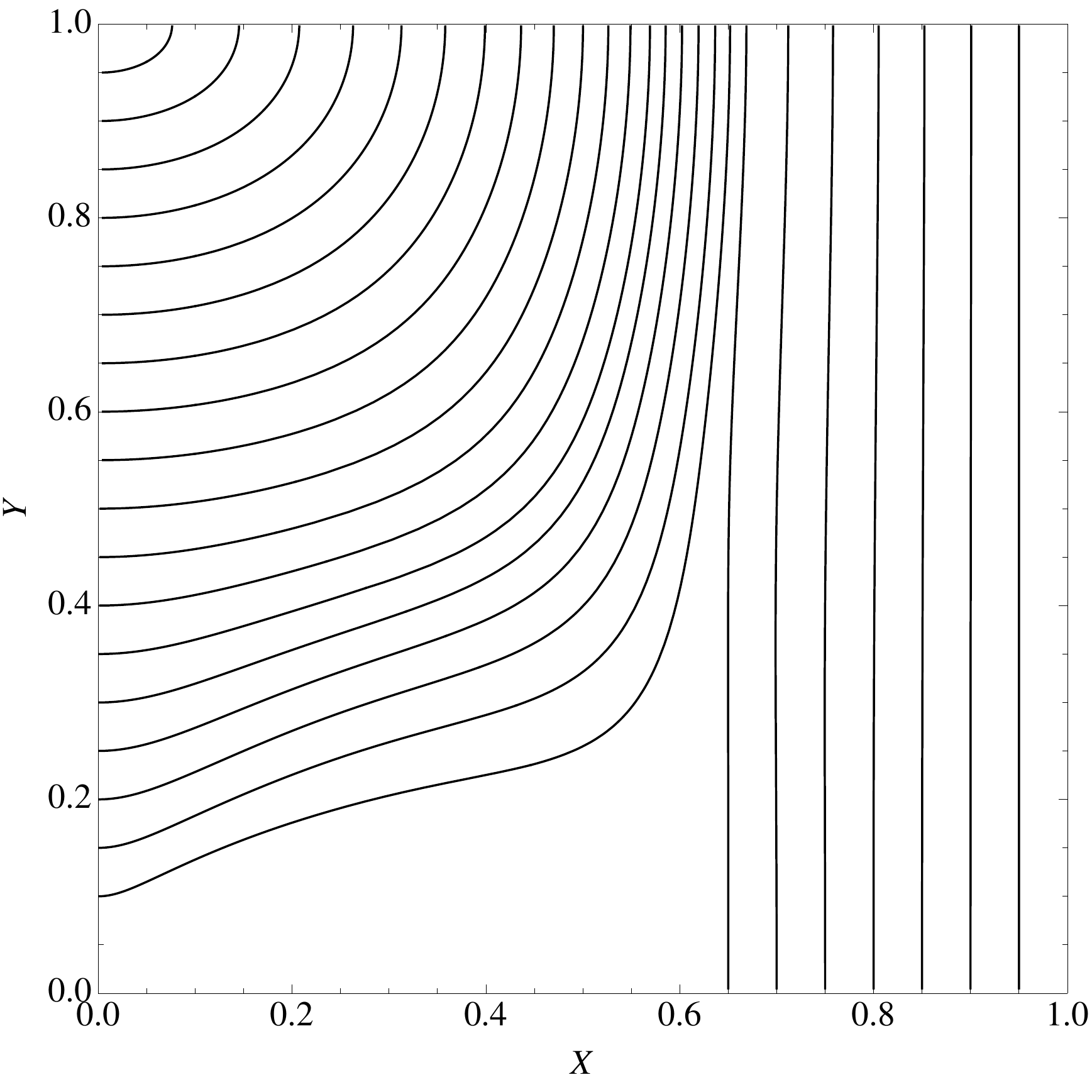}
\end{center}
\caption{Minimal surfaces in the background of a plasma ball of temperature $\beta/\beta_\textrm{crit}=0.78$. The topology of the minimal surfaces changes as we vary the size of the entangling region.}
\label{fig:entanglement}
\end{figure}

In Fig. \ref{fig:entanglement} we show families of minimal surfaces for a plasma ball in 5D with temperature $\beta/\beta_\textrm{crit}=0.78$. For this value of the inverse temperature, the size of the black hole along the IR bottom is $R_\mathrm{IR}/\ell =0.69$. Other values of $\beta$ give qualitatively similar results. Given our symmetry assumptions, these surfaces can be represented as a curve $Y(X)$ in the far region $(x,y)$ plane. As this plot shows, the topology of the minimal surfaces in the bulk changes as a function of the radius of the entangling region on the boundary. For $R_{EE}<R_{crit}$, the $\mathbf S^{d-3}$ of the minimal surface in the bulk shrinks to zero size somewhere on the axis of symmetry. This can be seen because the $X$ coordinate on this surface reaches zero for some finite value of $Y$. The larger the region $A$ is within the deconfined plasma, the closer the minimal surface gets to the horizon in the bulk. As this plot shows, as the minimal surfaces approach the horizon (the point $x=y=0$ in Fig. \ref{fig:entanglement}), the size of the entangling region on the boundary accumulates at a critical value $R_{crit}$.  For entangling regions of size $R_{EE}>R_{crit}$ the bulk topology of the minimal surface is different.  In this case,  it is the $\mathbf S^1_\mathrm{SS}$ of the minimal surface that shrinks to zero at the IR bottom whilst the $\mathbf S^{d-3}$ is always finite. This family of minimal surfaces correspond to the nearly vertical curves in Fig. \ref{fig:entanglement}, for which the $Y$ coordinate becomes zero at some point in $X$. Therefore, there is a phase transition among minimal surfaces at some $R_{EE}=R_{crit}$. The existence of this phase transition suggests that we can take this critical value of the entangling region as a measure of the size of the deconfined region, $R_{crit}\simeq R_{ball}$. For this particular value of $\beta/\beta_\textrm{crit}=0.78$, we estimate $R_{ball}/\ell\simeq 1.12$, which is compatible with the estimate that one gets from the temporal Wilson loop or the extent of the horizon along the IR bottom of the geometry.

\subsection{Dual stress tensor}
\label{subsec:stresstensor}
In this subsection we study the main features of the vacuum expectation value of the stress tensor of the plasma balls. Having found the plasma ball solutions numerically, we extract the stress tensor using the standard holographic renormalization techniques \cite{deHaro:2000xn}.
       
Using the prescription of \cite{deHaro:2000xn}, the resulting stress tensor is normalized such that it vanishes when the bulk geometry is AdS. This results in the AdS soliton having a non-zero vacuum expectation value which can be interpreted as Casimir energy. Since the AdS soliton is the lowest energy state obeying our boundary conditions \cite{Horowitz:1998ha}, in the following we will subtract the stress tensor of the soliton, so that the subtracted stress tensor vanishes on the AdS soliton background.  We find,
\begin{equation}
\begin{aligned}
\langle T_{\mu\nu}^{sub}\rangle\,dx^\mu\,dx^\nu = c_\textrm{eff}\,d\,\bigg[&-\frac{t^{(d)}(x)}{T_0}\,d\tau^2+s^{(d)}(x)\,d\theta^2+\frac{k_x^2\,x^2\,r^{(d)}(x)}{(1-x^2)^2}\,d\Omega_{(d-3)}^2\\
&+\frac{k_x^2(1+x^2)^2}{(1-x^2)^4}\left(\frac{t^{(d)}(x)}{T_0}-s^{(d)}(x)-(d-3)\,r^{(d)}(x)\right)\,dx^2\bigg]\,.
\end{aligned}
\label{eqn:stresstensor}
\end{equation}       
where $c_\textrm{eff}$ is the effective central charge \cite{Marolf:2013ioa},
\begin{equation}
c_\textrm{eff}=\frac{\ell^{d+1}}{16\,\pi\,G^{d+1}_N}\,.
\end{equation}
Note that \eqref{eqn:stresstensor} is manifestly traceless. As usual, this follows from solving the bulk equations of motion near the boundary. In \eqref{eqn:stresstensor} the functions $t^{(d)}(x)$, $s^{(d)}(x)$ and $r^{(d)}(x)$ denote the coefficients of the terms appearing at order $d$ in the near boundary expansion of the functions in the far-region metric \eqref{eqn:far} in our working coordinates,
\begin{equation}
\begin{aligned}
&T(x,y)=T_0+(1-y^2)^d\,t^{(d)}(x)+\ldots\\
&S(x,y)=1+(1-y^2)^d\,s^{(d)}(x)+\ldots\\
&R(x,y)=1+(1-y^2)^d\,r^{(d)}(x)+\ldots
\end{aligned}
\end{equation}
We can extract these functions from our numerical solutions by fitting  our data to this near boundary behaviour. Of course, these functions are not all independent; conservation of the stress energy tensor imposes one differential constraint among them. As is well-known, the conservation of the dual stress energy tensor is guaranteed by the bulk equations of motion. Therefore, we do not need to impose this constraint but we can use it to asses the numerical errors in stress energy tensor. In data shown below, we have estimated that the error is less than \%5.   

\begin{figure}[t]
\begin{center}
\includegraphics[scale=0.7]{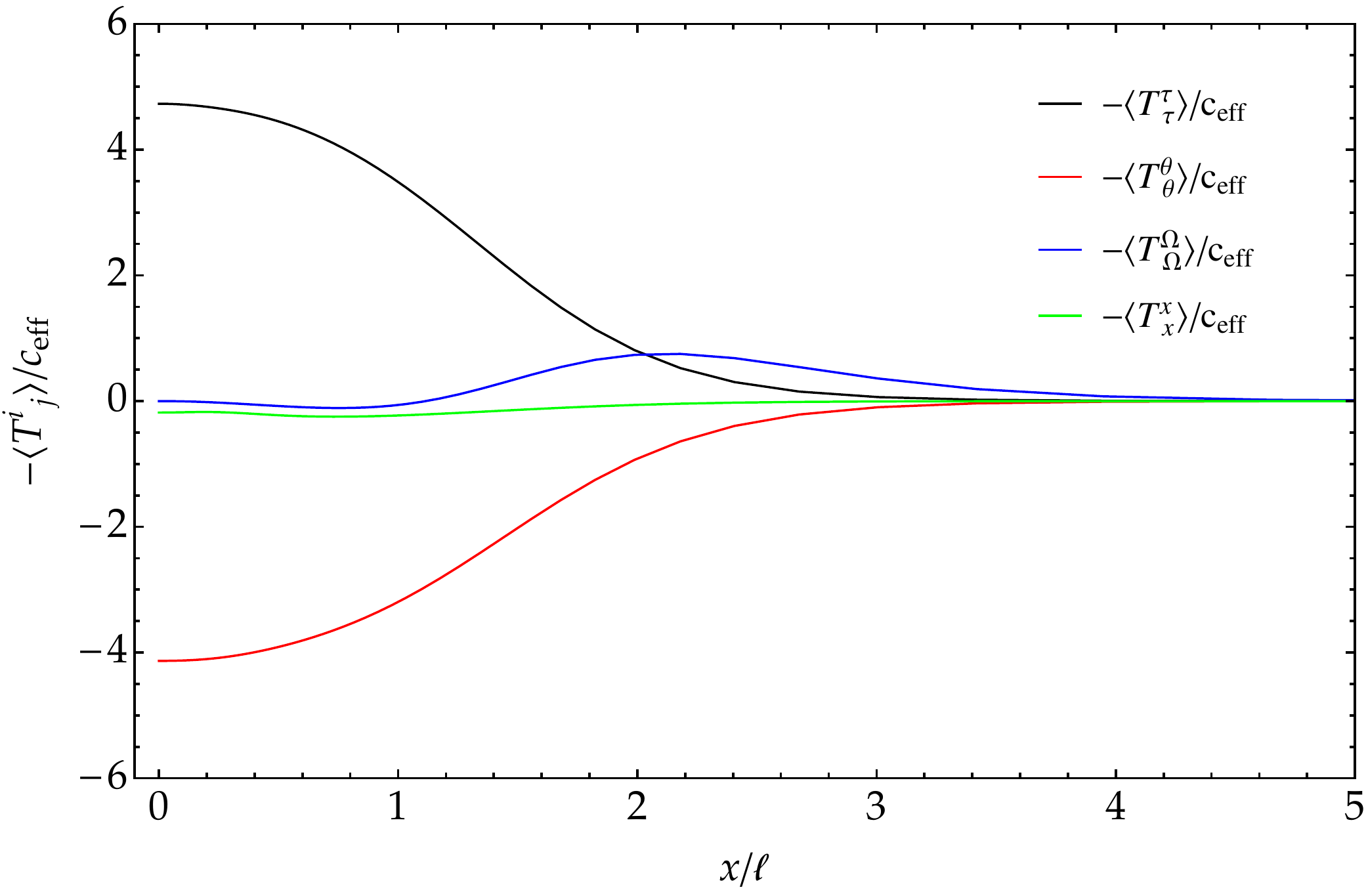}
\end{center}
\caption{Components of the subtracted stress tensor for a plasma ball with temperature $\beta/\beta_\textrm{crit} = 0.90$. This stress tensor shows some of the qualitative features of the stress tensor of the domain wall solution of \cite{Aharony:2005bm}. In particular, we find that the tension of the wall, measured by tangent components of the stress tensor, $-\langle T^\Omega_{~\Omega}\rangle$, is positive.}
\label{fig:stresstensor}
\end{figure}

In Fig. \ref{fig:stresstensor} we show the various components of the stress tensor for a five-dimensional plasma ball with temperature $\beta/\beta_\textrm{crit} = 0.90$ as a function of the radial coordinate along the boundary, $x$. This particular plasma ball is already quite large, $R_\mathrm{IR}/\ell = 1.35$. Therefore, it is not surprising that the stress tensor for this particular plasma ball has features which are qualitatively similar to those of the stress tensor of the domain wall solution of \cite{Aharony:2005bm} (see also \cite{Marolf:2013ioa}). In particular, at the centre of the ball, $\langle T_{\tau\tau}^{sub}\rangle$ and $\langle T_{\theta\theta}^{sub}\rangle$ have a similar magnitude and opposite sign. Also, note that the components of the stress tensor along the sphere directions ({\it i.e.,} the tangential components) obey $\langle T^{sub}_{\Omega\Omega}\rangle<0$ throughout the boundary of the ball, which results in a positive tension.

We can asses how close our large plasma balls are to the domain wall solution by having a closer look at the tension of the ball. Ref.  \cite{Aharony:2005bm} computed the tension in the domain wall limit and found,
\begin{equation}
\mu_{d-1} = C_d\,\epsilon_c\,\left(\frac{L}{2\pi}\right)^2\,\qquad C_4=2.0\,,\qquad C_5=1.7\,,
\end{equation}
where the constants $C_{4,5}$ are determined numerically.  For very large plasma balls of size $\rho\gg L$,  the interior of the ball should be well-described by the homogeneous deconfined plasma phase, and in the vicity of their edge these solutions should reduce to a domain wall. One can infer the behaviour of such large balls by balancing the pressure of the internal deconfined phase with the tension of the wall. One gets,
\begin{equation}
P=(d-3)\frac{\mu}{\rho}\,.
\label{eqn:pressure}
\end{equation}
Since the pressure $P$ of the deconfined phase as a function of the temperature is known, one can use \eqref{eqn:pressure} to obtain the corrections to temperature above the deconfinement transition as a function of the size $\rho$ of the ball \cite{Marolf:2013ioa},
\begin{equation}
T=\frac{1}{L}\left(1+\frac{(d-3)\,C_d}{(2\pi)^2}\frac{L}{\rho}+O\left(\textstyle{\frac{L^2}{\rho^2}}\right)\right)\,.
\end{equation}
Therefore, using this relation, we can independently estimate the values of $C_{4}$ and $C_{5}$ by considering data from our largest plasma balls and compare it to the values of obtained in \cite{Aharony:2005bm}. Our best estimates of these parameters are
\begin{equation}
C_4= 1.9\,,\qquad C_5 = 1.8\,.
\end{equation}

\section{Summary and conclusions}
\label{sec:summary}

In this paper we have constructed, numerically, localized spherical black holes solutions in 5D and 6D in the confining background of the AdS soliton. From the dual CFT perspective, our solutions correspond to localized balls of deconfined plasma surrounded by the confining vacuum. These solutions are parametrized by the temperature and they exist above the deconfinement temperature.

For temperatures much higher than the deconfinement temperature, these black holes are small (compared to the radius of AdS) and approximately spherically symmetric. On the other hand, as we approach the deconfinement temperature, the extent of the horizon along the IR bottom of the geometry grows without bound. Therefore, these large black holes look like pancakes. In the region near the edges, these large black holes look like domain walls, similar to the one constructed in \cite{Aharony:2005bm}, but at a different temperature. On the other hand, in the region far from the edges, the geometry of the space-time is well-approximated by the of a homogeneous black brane at the corresponding temperature.  As \cite{Aharony:2005bm} predicted, and we have confirmed in this paper, these localized black holes have the rather unusual property that in the infinite mass limit they have a constant temperature, namely the deconfinement temperature.

The black holes that we have constructed are thermodynamically unstable since they have negative specific heat, as they become cooler as their mass increases. At the deconfinement temperature their mass should become infinite. We have computed the spectrum of the Lichnerowicz operator around our solutions and we find that it has a single negative mode and no zero modes seem to appear along the branch of solutions. This suggests that these black holes  are presumably classically stable under perturbations that preserve the spatial symmetries of these backgrounds. This negative mode becomes smaller (in absolute value) as we approach the deconfinement temperature; this makes it computationally hard to construct very large black holes since the operator that we have to invert in every Newton step has a near-zero mode. It would be interesting to perform a detailed analysis of the quasi-normal mode spectrum of these black holes since it could provide interesting information of the near-equilibrium plasma physics in confining backgrounds. We will leave this for future work.  However, in the full quantum theory these black holes should be unstable to evaporate through the emission of Hawking radiation.

Ref. \cite{Lahiri:2007ae} constructed solutions to the Navier-Stokes equations that describe the plasma duals of rotating black holes and black rings in 5D and 6D SS--AdS. These results led to a prediction for the phase diagram of the dual black holes which exhibits some notorious differences with the asymptotically flat case; in particular, black rings should not be able to carry an arbitrarily large angular momentum per unit mass. In light of these results and predictions,  it would be very interesting to construct rotating black holes and black rings in SS--AdS.

In this paper we have computed temporal Wilson loops in the background of the plasma balls and we have verified that indeed their expectation value allows us to estimate the size of the deconfined region. We have also computed the entanglement entropy for certain spherically symmetric regions in the non-compact directions along the boundary. We have found that  the minimal surfaces in the bulk that compute the entanglement entropy undergo a topology change transition when the radial extent along the boundary of the entangling region is of the size of the deconfined region. Ref. \cite{Klebanov:2007ws} considered the entanglement entropy for strips of a certain width in confining backgrounds. They found a phase transition in the entanglement entropy qualitatively similar to the confinement/deconfinement phase transition. Our backgrounds contain a finite region of deconfined plasma ({\it i.e.,} the plasma ball) so it may seem that the phase transition that we have found is not obviously related to that found in \cite{Klebanov:2007ws}. It would be interesting to investigate this further. More recently, \cite{Kol:2014nqa} have compared Wilson loops and entanglement entropy in holography as probes of confinement. In this paper, the authors considered  a similar set up as \cite{Klebanov:2007ws} and they establish some criteria that the confining backgrounds should satisfy in order for the entanglement entropy to exhibit the aforementioned phase transition. It would be interesting to extend their criteria to the kinds of backgrounds and surfaces that we consider in this paper. 

Both Wilson loops and the entanglement entropy are observables that are computed from a minimization problem. Our solutions have two natural length scales, namely those set by the temperature and the mass gap. Therefore, it seems reasonable that these observables are sensitive to the ratio between these scales.\footnote{We are indebted to Carlos N\'u\~{n}ez for explaining this to us.} A qualitatively similar feature was observed for Wilson loops in backgrounds which exhibit ``walking" dynamics, which also have two length scales \cite{Nunez:2009da}.

\subsection*{Acknowledgements}
 PF would like to thank J. Lucietti for early collaboration in this project. We would also like to thank H. Reall, E. Perlmutter, R. Emparan, D. Mateos, A. Buchel, L. Lehner, R. Myers, C. N\'u\~nez, S. Minwalla, and especially T. Wiseman for discussions. PF and ST are supported by the European Research Council grant no. ERC-2011-StG279363-HiDGR. PF is also supported by the Stephen Hawking Advanced Research Fellowship from  Centre for Theoretical Cosmology, University of Cambridge.

\appendix

\section{Plasma walls}
\label{sec:plasmawalls}
With some minor modifications we can also construct finite energy density excitations about the AdS soliton background. These black holes have horizons with spatial topology $\mathbb R^{d-3}\times\mathbf S^2$ and hence they are extended in the flat directions of the bulk geometry.  From the perspective of the dual CFT, these black holes are localized in one of the $\mathbb R^{d-2}$ directions and extended in the remaining ones; as in the plasma balls case, they do not extend along the $\mathbf S^1_\mathrm{SS}$. Note that in the case of $d=4$ ({\it i.e.,} five bulk dimensions), from the boundary perspective these configurations will look like plasma strips;  in the higher dimensional cases they will look like plasma walls and therefore we shall refer to them as such.

For simplicity we will restrict ourselves to configurations which respect the $\mathbb R^{d-3}\times U(1)_\mathrm{SS}$ of the spatial flat boundary metric. Note that these black holes should be expected to suffer from Gregory-Laflamme type of instabilities along the extended directions of the horizon; since in our construction we will (consistently) neglect the dependency on these extended flat directions we will not ``see" the zero modes signalling the onset of these instabilities. Whilst studying these instabilities is certainly of interest, it is beyond the scope of this paper.

The plasma wall solutions can be simply constructed by replacing the $(d-3)$-sphere appearing in the finite size plasma balls by a $(d-3)$-plane. Doing so, results in a  far region ansatz of the form,
\begin{equation}
\begin{aligned}
ds^2_{Far}=&~\frac{1}{(1-y^2)^2}\Bigg(T\,d\tau^2+\frac{4\,y^2\,\bar f(y)\,S}{d^2}\,d\theta^2+R\,d\mathbf{x}_{(d-3)}^2\\
&\hspace{2cm}+\frac{k_x^2(1+x^2)^2\,A}{(1-x^2)^4}\,dx^2+\frac{4\,B}{\bar f(y)}\,dy^2-\frac{2\,k_x\,(1+x^2)\,F}{(1-x^2)^2(1-y^2)}\,dx\,dy\Bigg)\,.
\end{aligned}
\label{eqn:Wfar}
\end{equation} 
The far region coordinates $(x,y)$ have the same ranges as in \S\ref{subsubsec:BCs}. The only difference is that now $x=0$ is a reflection plane. Therefore, on this boundary the various functions $\{T,~R,~S,~A,~B,~F\}$ satisfy the same boundary conditions that we have outlined in \S\ref{subsubsec:BCs}, except that now we do {\it not} have to impose the absence of conical singularities. At the other boundaries of the far region chart, the functions satisfy the same boundary conditions as in \S\ref{subsubsec:BCs}. 

Similarly,  the near region metric ansatz becomes, 
\begin{equation}
\begin{aligned}
ds^2_{Near} =&~\frac{1}{(1-r^2)^2}\bigg[r^2\,g(r)\,T'\,d\tau^2 + r_0^2\Big(R'\,d\mathbf{x}_{(d-3)}^2+a^2(2-a^2)\,S'\,d\theta^2\Big) \\
&\hspace{2cm}
+\frac{4\,r_0^2\,A'}{g(r)}\,dr^2 + \frac{4\,r_0^2\,B'}{2-a^2}\,da^2 + 2\,r\,F'\,dr\,da\bigg]\,.
\end{aligned}
\label{eqn:Wnear}
\end{equation}   
The near region coordinates $(r,a)$ have the same ranges as in \S\ref{subsubsec:BCs} but now $a=1$ is a reflection plane. There, the functions $\{T',~R',~S',~A',~B',~F'\}$ satisfy the same boundary conditions as before ({\it i.e.,} Neumann) except that now we do not impose the absence of conical singularities. The remaining boundaries are treated as in the plasma balls case.  The rest of the numerical construction proceeds in exactly the same way as before.

The behaviour of the area (density) as a function of the extent of the wall along the IR bottom of the geometry is qualitatively similar to the plasma balls case. In particular, we find at it diverges at the deconfinement temperature. In the opposite high temperature limit, the plasma walls tend to asymptotically flat black branes.  Within this symmetry class of metrics we do not expect any qualitatively new physics; in particular, the temporal Wilson loops and the entanglement entropy should behave in qualitatively similar manner as in the plasma balls case. We leave a detailed study of the physical properties of these solutions for future work.

\section{Convergence tests}
\label{sec:convergence}

\begin{figure}[t]
\begin{center}
\includegraphics[width=15cm]{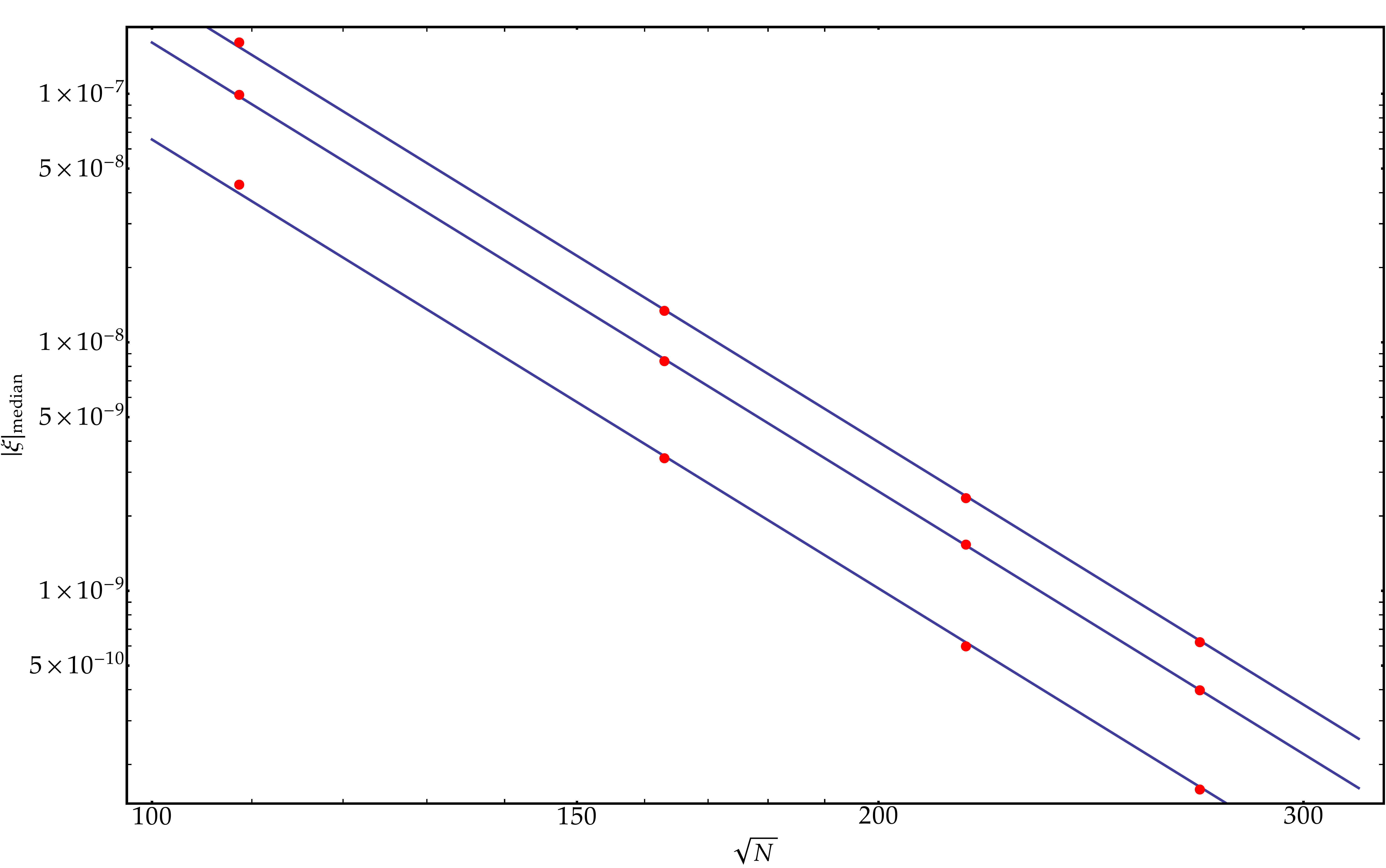}
\end{center}
\caption{The decay of error in three representative solutions (from top down) $\beta / \beta_\mathrm{crit} = 0.5, 0.7, 0.6$ as the grid resolution is increased. Here the error is estimated with the median value of $1 + R / 20$, and $N$ is the total number of grid points used. Note that $\sqrt{N}$ scales like $h^{-1}$ where $h$ is the typical grid spacing. Blue lines show the best-fitting $h^{-6}$ functions.}
\label{fig:err}
\end{figure}

In this appendix, we inspect the behaviour of the error in our numerical solutions as the grid resolution is varied. We present this in Fig. \ref{fig:err} as a plot of the median magnitude of the scalar $1 + R / 20$ as a function of the square root of the number of grid points $N$, which scales as the inverse of the typical grid spacing $h$. Since we are using sixth-order finite differences to solve the equations, if the solution is at least six-times differentiable then the error should decay like $h^{-6}$ as the continuum limit is approached. Indeed, in the three sample cases shown this is precisely what we observe, which provides reassurance that our numerical solution is a good approximation to the actual solution to the continuous problem.

\bibliographystyle{utphys}
\bibliography{plasmaballs}

\end{document}